\documentclass[11pt,a4paper]{article}

\usepackage{amsmath,amssymb,wasysym,cite,epsfig,graphicx}

\begin{document}
	
\title{Shadow of axisymmetric, stationary and asymptotically flat black holes in the presence of plasma}

\author{Javier Bad\'ia$^{1, 2}$\thanks{e-mail: jbadia@iafe.uba.ar} and Ernesto F. Eiroa$^{1}$\thanks{e-mail: eiroa@iafe.uba.ar}\\
	{\small $^1$ Instituto de Astronom\'{\i}a y F\'{\i}sica del Espacio (IAFE, CONICET-UBA),}\\
	{\small Casilla de Correo 67, Sucursal 28, 1428, Buenos Aires, Argentina}\\
	{\small $^2$ Departamento de F\'{\i}sica, Facultad de Ciencias Exactas y Naturales,} \\ 
	{\small Universidad de Buenos Aires, Ciudad Universitaria Pabell\'on I, 1428, Buenos Aires, Argentina}}
\date{}

\maketitle

\begin{abstract}
We study the shadow produced by a class of rotating black holes surrounded by plasma. The metric for these black holes arises by applying the Newman-Janis algorithm to a family of spherically symmetric spacetimes, which includes several well known geometries as special cases. We derive a general expression for the shape of the shadow in the case that the plasma frequency leads to a separable Hamilton-Jacobi equation for light. We present two examples in which we obtain the shadow contours and the observables resulting from them. In one, we analyze Kerr-Newman-like geometries, including braneworld and Horndeski gravity black holes,  while in the other, we consider scalar-tensor 4D Einstein-Gauss-Bonnet gravity spacetimes. In both cases, we find that the presence of plasma leads to a smaller and less deformed shadow.
\end{abstract}

\section{Introduction}\label{intro}

Two years ago, the international collaboration Event Horizon Telescope (EHT) announced the first reconstructed image \cite{eht19} of the supermassive black hole at the center of the giant elliptical galaxy M87  \cite{broderick15}, which shows the shadow surrounded by the light coming from the accretion disk around M87*; the observed bright emission ring has a diameter of $42 \pm 3$ $\mu $as. This telescope consists of a very long baseline interferometry (VLBI) array of instruments spread over the Earth, operating in millimeter radio waves (230 GHz). The shadow of the supermassive black hole Sgr A* at the center of the Milky Way \cite{gillessen17} is also of interest for this kind of observation, but its imaging has not been fully achieved yet. The shadow or apparent shape of a black hole --as seen by a far away observer-- is a consequence of the particular behavior of the photons in the vicinity of these compact objects. In the case of a non-rotating black hole it has the shape of a circle, while rotating ones present a deformation that increases with the spin \cite{bardeen,chandra}. Many articles have been published on this topic in the years previous to the EHT discovery, both in Einstein theory \cite{luminet,shape,tsupko17,other18} and in modified gravity \cite{sombrasvarios,braneworld,sombrasvarios2,other16,tsukamoto18}. 
In general relativity, the size and the shape of the shadow depend on the mass, the angular momentum, and the inclination angle of the black hole, along with other variables that can appear due to the presence of matter or fields; while in modified gravity they can also depend on other parameters related to the particular theory adopted. There has been a surge in the number of articles published on this topic since the EHT discovery; we can mention Refs. \cite{new19rg,new19at,ghosh20,new20rg,new20at,badia20} among them. For an up to date review of analytical studies, see Ref. \cite{ps21}. Interesting discussions about the physical nature of the black hole photon ring and the shadow have recently appeared in the literature \cite{topics}. The study of black hole shadows can be a useful tool for a better understanding of astrophysical black holes at the centers of galaxies. The measurement of black hole shadows has been proposed for testing general relativity in the strong field regime, allowing for a comparison with other theories of gravity \cite{tests}. An improvement in the observations by the EHT is expected in the coming years, which would lead to a more detailed picture of M87* and also to the acquisition of images of other nearby supermassive black holes \cite{observ,millimetron,ehi-fromm21}.

The presence of plasma surrounding an astrophysical object results in a change of the trajectories of light rays with respect to those in a vacuum background. But plasma is also a dispersive medium, so photons with different frequencies follow distinct trajectories; as a consequence, the optical properties are chromatic. In the context of geometrical optics, the plasma can be considered as a dispersive medium characterized by an index of refraction depending on the frequency. One particular aspect that has been analyzed in recent years is the influence of plasma on the shadows of black holes, both in the spherically symmetric  \cite{perlick15} and in the rotating \cite{atamurotov15,perlick17,kogan17,plasma-other,junior20} cases. In this article, we investigate the shadow cast by a class of rotating black holes surrounded by plasma, which results from applying the Newman-Janis algorithm to spherically symmetric spacetimes. We obtain a general expression for the contour of the shadow when the plasma frequency leads to a separable Hamilton-Jacobi equation. We present examples corresponding to Kerr-Newman-like and scalar-tensor 4D Einstein-Gauss-Bonnet theory black holes, for which we introduce and calculate the observables. The paper is organized as follows. In Sec. \ref{HJeq}, we separate the Hamilton-Jacobi equation in the case of photons in plasma. In Sec. \ref{photons}, we find the contour of the shadow. In Sec. \ref{examples}, we show the examples. Finally, in Sec. \ref{conclu}, we discuss the main results. We adopt units such that $G=c=\hbar=1$.

\section{Hamilton-Jacobi equation for light rays in a plasma}\label{HJeq}

The motion of photons in a pressureless, nonmagnetized plasma is governed by the Hamiltonian \cite{synge60,perlick00}
\begin{equation}\label{eq:hamiltoniano}
	\mathcal{H}(x,p) = \frac{1}{2} \left( g^{\mu\nu}(x) p_\mu p_\nu + \omega_p(x)^2 \right),
\end{equation}
where $g^{\mu\nu}$ is the inverse metric, $x=(x^0,x^1,x^2,x^3)$ denotes the spacetime coordinates, $p=(p^0,p^1,p^2,p^3)$ represents the conjugate momentum, and $\omega_p$ is the plasma electron frequency, given by
\begin{equation}\label{eq:plasma-densidad}
	\omega_p(x)^2 = \frac{4\pi e^2}{m_e} N_e(x),
\end{equation}
with $e$ and $m_e$ being the electron charge and mass, respectively, and $N_e$ the electron number density. The vacuum case corresponds to $\omega_p(x)=0$. The plasma is a dispersive medium, with a squared index of refraction \cite{perlick17}
\begin{equation}\label{eq:plasma-n}
	n(x,\omega)^2 = 1 - \frac{\omega_p(x)^2}{\omega(x)^2},
\end{equation}
that depends on the photon frequency $\omega(x)$ with respect to the plasma. Light propagation in this medium is only possible  if 
\begin{equation}\label{eq:plasma-ineq}
	\omega(x) \ge \omega_p(x),
\end{equation}
which guarantees a real and non-negative index of refraction.

In what follows, we assume that both the metric and the plasma frequency are stationary and axisymmetric, so that in the Boyer-Lindquist coordinates $(t,r,\theta,\varphi)$ neither the components $g_{\mu\nu}$ nor $\omega_p$ can depend on $t$ and $\varphi$. Since for a light ray $\mathcal{H}(x,p) = 0$, the form of the Hamiltonian introduced in Eq. \eqref{eq:hamiltoniano} implies that light in the presence of plasma follows timelike worldlines for the metric $g_{\mu\nu}$. A given $\omega_p$ also fixes the normalization of the curve parameter, with the Cartesian components of the four-momentum having units of frequency. For this reason, we will follow Ref. \cite{perlick17} and use the name $\omega_0 \equiv -p_t$ for the conserved photon energy $E=\omega_0$; the quantity $\omega_0$ is the photon frequency measured by an observer at infinity\footnote{Note, however, that there is a difference from \cite{perlick17} in our definition, in which $\omega_0$ is positive for a future directed vector $p^\mu$, since we are interested in future directed rather than past directed rays.}. Another consequence of the presence of plasma is the existence of forbidden regions, where light rays cannot exist. 
When the frequency $\omega_0$ is small enough compared with the plasma frequency  $\omega_p(x)$, the propagation of light in the plasma is not possible. The analysis carried out in Ref. \cite{perlick17} for the Kerr spacetime can be straightforwardly generalized to an arbitrary stationary, axisymmetric and asymptotically flat metric, with the result that
\begin{equation}\label{eq:prohibida}
	\omega_0^2 \geq - g_{tt} \omega_p^2(r,\theta)
\end{equation}
is a necessary and sufficient condition for at least one light ray with a given frequency $\omega_0$ to exist at a given spacetime point. 

We adopt the G\"urses-G\"ursey geometry \cite{gurses75}, defined by the line element that in Boyer-Lindquist coordinates reads  
\begin{equation}\label{eq:metrica}
	ds^2 = - \frac{\rho^2\Delta}{\Sigma} dt^2 + \frac{\Sigma \sin^2\theta}{\rho^2} \left[d\varphi - \frac{2a m(r) r}{\Sigma} dt\right]^2 + \frac{\rho^2}{\Delta} dr^2  + \rho^2 d\theta^2,
\end{equation}
where
\begin{align}
	\rho^2 &= r^2 + a^2 \cos^2\theta, \\
	\Delta &= r^2 - 2m(r)r + a^2, \\
	\Sigma &= (r^2+a^2)^2 - a^2 \Delta \sin^2\theta.
\end{align}
Here $m(r)$ is a function that approaches the mass $M$ of the black hole as $r \to \infty$, and $a = J/M$ is the angular momentum per unit mass of the black hole. This spacetime arises in different contexts; it was first derived from a metric of the Kerr-Schild class by using the Newman-Janis complex transformation \cite{gurses75} and it was subsequently reobtained \cite{bambi13} by applying the Newman-Janis algorithm to a spherically symmetric seed metric of the form\footnote{An extra dependence on the angular coordinate $\theta$ is introduced in Ref. \cite{bambi13}, i.e. $m(r,\theta)$, which is not relevant for our work.}
\begin{equation}
	ds^2 = - \left(1 - \frac{2m(r)}{r}\right) dt^2 + \left(1 - \frac{2m(r)}{r}\right)^{-1} dr^2 + r^2 (d\theta^2 + \sin^2\theta\, d\varphi^2).
\end{equation}
The Newman-Janis algorithm should be applied with care outside general relativity \cite{hansen13}, because the static and the resulting rotating solutions may correspond to different energy-momentum tensors. The shadows of black holes defined by the metric \eqref{eq:metrica} were recently studied \cite{tsukamoto18} without the presence of plasma. In order that the geometry \eqref{eq:metrica} represents a rotating black hole, we assume that the equation $\Delta(r) = 0$ has one or more positive solutions, the largest of which corresponds to the event horizon. It is easily seen that this spacetime is asymptotically flat, stationary, and axisymmetric, and therefore the quantities $p_t = -\omega_0$ and $p_\varphi$ are conserved along the geodesics of photons; $p_\varphi$ is the $z$ component of the angular momentum. Since $\omega_p$ is a function of the $r$ and $\theta$ coordinates only, these quantities are still conserved in the presence of plasma. A third constant of motion is $\mathcal{H} =0$.

To find one more conserved quantity and bring the equations of motion to first-order form, we write down the Hamilton-Jacobi equation for light rays
\begin{equation}\label{eq:hamilton-jacobi}
	\mathcal{H}\left(x, \frac{\partial S}{\partial x}\right) = 0,
\end{equation}
and we attempt to separate variables with the ansatz
\begin{equation}
	S = - \omega_0 t + p_\varphi \varphi + S_r(r) + S_\theta(\theta).
\end{equation}
Substituting into Eq. \eqref{eq:hamilton-jacobi}, we arrive at
\begin{equation}\label{eq:hj-2}
	\Delta (S_r')^2 - \frac{1}{\Delta} \left[(r^2+a^2)^2 \omega_0^2 - 4a m(r) r \omega_0 p_\varphi + a^2 p_\varphi^2\right] 
	+ (S_\theta')^2 + a^2 \omega_0^2 \sin^2\theta + \frac{p_\varphi^2}{\sin^2\theta} + \rho^2 \omega_p^2 = 0,
\end{equation}
where the prime denotes the derivative with respect to $r$ or $\theta$, as appropriate. It can be seen, as previously shown for the Kerr spacetime \cite{perlick17}, that this equation is separable if and only if the plasma frequency can be written in the form
\begin{equation}\label{eq:plasma-sep}
	\omega_p^2 = \frac{f_r(r) + f_\theta(\theta)}{\rho^2},
\end{equation}
with $f_r$ and $f_\theta$ being functions of their respective coordinates. We take the plasma frequency to be of this form from now on. By substituting this expression for $\omega_p$ into the Hamilton-Jacobi equation \eqref{eq:hj-2}, we can separate it as
\begin{equation}
	(S_\theta')^2 + \left(a \omega_0 \sin\theta - \frac{p_\varphi}{\sin\theta}\right)^2 + f_\theta = -\Delta (S_r')^2 + \frac{1}{\Delta} \left[(r^2+a^2)\omega_0 - a p_\varphi\right]^2 - f_r.
\end{equation}
Since the left hand side is a function only of $\theta$ and the right hand side a function only of $r$, they must both be equal to a constant $\mathcal{K}$. For future convenience, we use instead the Carter constant \cite{carter68}, defined by $\mathcal{Q} = \mathcal{K} - (p_\varphi - a\omega_0)^2$, as the fourth conserved quantity of motion.

From Hamilton equations, the derivatives of the $t$ and $\varphi$ coordinates can be found from $\dot{x}^\mu = p^\mu = g^{\mu\nu}p_\nu$, where the dot represents the derivative with respect to a curve parameter $\lambda$, which does not have a direct physical meaning in the presence of plasma \cite{perlick17}; in vacuum $\lambda$ is the affine parameter. By setting the covariant momenta equal to the derivatives of $S$, i.e. $p_\nu = \partial S/dx^\nu $, the equations of motion can then be brought to the first-order ones:
\begin{align}
	\rho^2 \dot{t} &= \frac{r^2 + a^2}{\Delta}P(r) - a(a\omega_0\sin^2\theta - p_\varphi), \label{eq:t-dot} \\
	\rho^2 \dot{r} &= \pm \sqrt{R(r)}, \label{eq:r-dot} \\
	\rho^2 \dot{\theta} &= \pm \sqrt{\Theta(\theta)},\label{eq:theta-dot} \\
	\rho^2 \dot{\varphi} &= \frac{a}{\Delta}P(r) - a\omega_0 + \frac{p_\varphi}{\sin^2\theta}, \label{eq:phi-dot}
\end{align}
where
\begin{align}
	R(r) &= P(r)^2 - \Delta [\mathcal{Q} + (p_\varphi - a \omega_0)^2 + f_r], \\
	\Theta(\theta) &= \mathcal{Q} + \cos^2\theta \left(a^2 \omega_0^2 - \frac{p_\varphi^2}{\sin^2\theta}\right) - f_\theta, \\
	P(r) &=(r^2 + a^2)\omega_0 - a p_\varphi.
\end{align}
We have finally arrived at the equations of motion for photons in a spacetime of the form given by Eq. \eqref{eq:metrica} in the presence of plasma. Note that the dependence with $m(r)$ comes through $\Delta$. After a suitable identification of the metric functions, it is straightforward to verify that our results agree with those obtained in Ref. \cite{junior20}\footnote{In this related work, the Hamilton-Jacobi equation is separated in the presence of plasma, but is not explored how this affects the shadow.}.

\section{Photon orbits and the shadow of the black hole}\label{photons}

We are now interested in the spherical photon orbits, defined as the geodesics that stay at a constant value of $r$. These trajectories will serve as the limiting case for the rays that form the boundary of the black hole shadow.

\subsection{Spherical photon orbits}\label{orbits}

Finding the trajectories with constant $r$ requires, by the radial equation of motion, to obtain the solutions of the simultaneous equations $R(r) = R'(r) = 0$. We assume that such solutions exist and that they are unstable, satisfying $R''(r) > 0$. We follow the standard method, described for example in Ref. \cite{tsukamoto18}, for the calculation of the constants of motion $p_\varphi$ and $\mathcal{Q}$ in terms of the radius $r$ of the spherical photon orbit. From the equation $R(r) = 0$ we can easily solve for $\mathcal{Q}$:
\begin{equation}\label{eq:eta-xi}
	\mathcal{Q} + (p_\varphi-a \omega_0)^2 + f_r = \frac{\left[(r^2+a^2)\omega_0 - a p_\varphi \right]^2}{\Delta}.
\end{equation}
Then, substituting this result into $R'(r) = 0$, we have
\begin{equation}\label{eq:rp-xi}
	R' = 4\omega_0 r \left[(r^2+a^2)\omega_0 - a p_\varphi \right] -\frac{\Delta'}{\Delta} \left[ (r^2+a^2)\omega_0 - a p_\varphi \right]^2 - \Delta f_r' = 0,
\end{equation}
which is a quadratic equation for $p_\varphi$, with the solution
\begin{equation}\label{eq:pphi-c}
	p_\varphi = \frac{\omega_0}{a}\left[ r^2 + a^2 - \frac{2r\Delta}{\Delta'} \left(1 \pm \sqrt{1 - \frac{\Delta' f_r'}{4 \omega_0^2 r^2}} \right)  \right],
\end{equation}
where $\Delta' = 2[r-m'(r)r-m(r)]$. Finally, Eq. \eqref{eq:pphi-c} has to be substituted into Eq. \eqref{eq:eta-xi} to solve for $\mathcal{Q}$ as a function of $r$. A possibly helpful intermediate step is to use Eq. \eqref{eq:rp-xi} to arrive at the relation
\begin{equation}
	\frac{\Delta'}{\Delta} a^2 (p_\varphi - a\omega_0)^2 = 2a \omega_0 r \left(\frac{r \Delta'}{\Delta} - 2\right) (p_\varphi-a\omega_0) + 4 \omega_0^2 r^3 - \frac{\Delta'}{\Delta} \omega_0^2 r^4 - \Delta f_r',
\end{equation}
which gives the square of $p_\varphi-a\omega_0$. Putting all together, we get
\begin{equation}\label{eq:q-c}
	\mathcal{Q} = - \frac{\omega_0^2 r^4}{a^2} + \frac{4\omega_0^2 r^2 \Delta}{a^2\Delta'} \left[r - \frac{2}{\Delta'}\left(\Delta - a^2\right)\right] \left(1 \pm \sqrt{1 - \frac{\Delta' f_r'}{4\omega_0^2 r^2}} \right) + \frac{\Delta f_r'}{\Delta' a^2} \left(\Delta - a^2\right) - f_r.
\end{equation}
For a radius $r$ and a frequency $\omega_0$, Eqs. \eqref{eq:pphi-c} and \eqref{eq:q-c} give the critical values of the conserved quantities $p_\varphi$ and $\mathcal{Q}$ associated with the corresponding spherical orbit. From Eq. \eqref{eq:theta-dot}, we see that any trajectory must have $\Theta \geq 0$. If for a fixed $r$ we then substitute the critical values of $p_\varphi$ and $\mathcal{Q}$ into the inequality $\Theta \geq 0$, we arrive at
\begin{equation}\label{eq:reg-fotones}
	\mathcal{Q} + \cos^2\theta \left( a^2 \omega_0^2 - \frac{p_\varphi^2}{\sin^2\theta}\right) - f_\theta(\theta) \geq 0,
\end{equation}
where $\mathcal{Q}$ and $p_\varphi$ are given by Eqs. \eqref{eq:pphi-c} and \eqref{eq:q-c}. The region defined by Eq. \eqref{eq:reg-fotones} is known as the photon region, and spherical photon orbits exist at values of $r$ and $\theta$ for which this inequality is satisfied. The range of possible radii in Eqs. \eqref{eq:pphi-c} and \eqref{eq:q-c} consists of those values of $r$ for which there exists at least one value of $\theta$ satisfying Eq. \eqref{eq:reg-fotones}.

\subsection{Black hole shadow}\label{shadow}

For a far away observer, the black hole shadow is the set of directions in the sky which, when propagated backwards in time, never reach infinity and instead cross the event horizon. Its boundary consists of those rays that asymptotically approach the spherical photon orbits of the spacetime, and which therefore have the same conserved quantities as them. In order to relate the directions in the sky with the constants of motion of the light ray, we take an observer at rest in the asymptotically flat region (large $r_\text{o}$) with an inclinaton angle $\theta_\text{o}$ from the spin axis of the black hole, and we construct the orthonormal tetrad 
\begin{align}
	\mathbf{e}_{\hat{t}} &= \partial_t, \\
	\mathbf{e}_{\hat{r}} &= \partial_r, \\
	\mathbf{e}_{\hat{\theta}} &= \frac{1}{r_\text{o}} \partial_\theta, \\
	\mathbf{e}_{\hat{\varphi}} &= \frac{1}{r_\text{o} \sin\theta_\text{o}} \partial_\varphi,
\end{align}
so that the corresponding components of the four-momentum for a photon with frequency $ \omega_0$ are given by
\begin{align}
	p^{\hat{t}} &= \omega_0, \\
	p^{\hat{r}} &= p^r, \\
	p^{\hat{\theta}} &= r_\text{o} p^\theta, \\
	p^{\hat{\varphi}} &= r_\text{o} \sin\theta_\text{o} p^\varphi = \frac{p_\varphi}{r_\text{o} \sin\theta_\text{o}}.
\end{align}
For the plasma model, we assume that 
\begin{equation}
	\lim_{r \to \infty} \omega_p(r, \theta) =0, 
\end{equation}
or equivalently
\begin{equation}
	\lim_{r \to \infty} \frac{f_r(r)}{r^2}=0,
\end{equation}
so photons propagate in vacuum far away from the black hole. We then adopt the celestial coordinates for an observer at infinity \cite{bardeen,chandra}:
\begin{gather}
	\alpha = - r_\text{o} \frac{p^{\hat{\varphi}}}{p^{\hat{t}}} \bigg|_{r_\text{o} \to \infty}, \\
	\beta = - r_\text{o} \frac{p^{\hat{\theta}}}{p^{\hat{t}}} \bigg|_{r_\text{o} \to \infty},
\end{gather}
and we insert the expression \eqref{eq:theta-dot} for $p^\theta$ as a function of the conserved quantities to finally arrive at
\begin{gather}
	\alpha = - \frac{p_\varphi}{\omega_0 \sin\theta_\text{o}}, \label{eq:alfa}\\
	\beta = \pm \frac{1}{\omega_0} \sqrt{\mathcal{Q} + \cos^2\theta_\text{o} \left(a^2 \omega_0^2 - \frac{p_\varphi^2}{\sin^2\theta_\text{o}}\right) - f_\theta(\theta_\text{o})}. \label{eq:beta}
\end{gather}
For a given $\omega_0$, the contour of the black hole shadow is described by a parametric curve $(\alpha(r), \beta(r))$, with $p_\varphi$ and $\mathcal{Q}$ given as functions of $r$ by Eqs. \eqref{eq:pphi-c} and \eqref{eq:q-c}, and $r$ bounded in the region $r_+ \leq r \leq r_-$, with $r_\pm$ being the values for which $\beta(r_\pm) = 0$. Comparing with Eq. \eqref{eq:reg-fotones}, this is the intersection of the photon region with the cone $\theta = \theta_\text{o}$. The directions of $\alpha$ and $\beta$ are, respectively, perpendicular and parallel to the spin of the black hole.

\subsection{Observables}

In order to characterize the black hole shadow, following our previous work \cite{badia20}, we calculate three observables: the area of the shadow, its oblateness, and the horizontal displacement of its centroid \cite{tsupko17,badia20,ghosh20}. The area is simply defined by
\begin{equation}
	A = 2 \int \beta\, d\alpha = 2 \int_{r_+}^{r_-} \beta(r) |\alpha'(r)|\, dr,
\end{equation}
with the factor $2$ arising from the up-down symmetry of the shadow, since the curve $(\alpha(r), \beta(r))$ only describes half of the contour because we are taking the plus sign in Eq. \eqref{eq:beta}. The oblateness is related to the deformation of the shadow as compared to a circle, and is defined as
\begin{equation}
	D = \frac{\Delta \alpha}{\Delta \beta},
\end{equation}
where $\Delta\alpha$ and $\Delta\beta$ are the horizontal and vertical extents of the shadow, respectively; the Kerr shadow has $D \le 1$, with $D=1$ for the limiting case of a circle. Finally, the horizontal coordinate of the centroid is given by
\begin{equation}
	\alpha_c = \frac{2}{A} \int \alpha \beta\, d\alpha = \frac{2}{A} \int_{r_+}^{r_-} \alpha(r) \beta(r) |\alpha'(r)|\, dr,
\end{equation}
with the same factor of $2$ as in the definition of the area.

\section{Examples}\label{examples}

In this section, we obtain the contour of the shadow for two black hole geometries by using the formalism described in Sec. \ref{photons} and we calculate the three observables defined there in order to characterize their size and shape, as functions of the parameters of each example. We adopt the well known case of dust that is at rest at infinity, first considered by Shapiro \cite{shapiro74}, as our model for the plasma. In the Kerr spacetime the mass density, and by Eq. \eqref{eq:plasma-densidad} the squared plasma frequency, go as $r^{-3/2}$, being independent of $\theta$ to a very good approximation\footnote{The density actually goes to a constant at infinity, with $\rho \sim r^{-3/2}$ only being true at distances below the capture radius of the black hole. We consider the density at infinity to be negligible, and take $\rho \propto r^{-3/2}$ everywhere.}. However, such a plasma distribution cannot be put into the separable form given by Eq. \eqref{eq:plasma-sep}; therefore, following Ref. \cite{perlick17}, we take the frequency to have an additional $\theta$ dependency by choosing
\begin{gather}
	f_r(r) = \omega_c^2 \sqrt{M^3 r}, \\
	f_\theta(\theta) = 0,
\end{gather}
so that
\begin{equation}\label{eq:plasma-shapiro}
	\omega_p^2 = \omega_c^2 \frac{\sqrt{M^3 r}}{r^2 + a^2\cos^2\theta},
\end{equation}
where $\omega_c$ is a constant and $M$ is the mass of the black hole. At large distances from the black hole, the metrics considered in this work approach the Kerr metric as long as we have $m(r) \to M$, so we expect that the Shapiro solution is still valid. At small distances, the particular metric may be quite different from the Kerr one, so we will work under the assumption that the plasma density given by Eq. \eqref{eq:plasma-shapiro} is not significantly affected. It is not difficult to see, by combining Eqs.  \eqref{eq:pphi-c} and \eqref{eq:q-c} with Eqs. \eqref{eq:alfa} and \eqref{eq:beta}, that the contour of the shadow for a given photon frequency $\omega_0$ is determined by the ratio $\omega_c/\omega_0$.

\subsection{Kerr-Newman-like black holes}

\begin{figure}[t!]
	\centering
	\includegraphics[width=\textwidth]{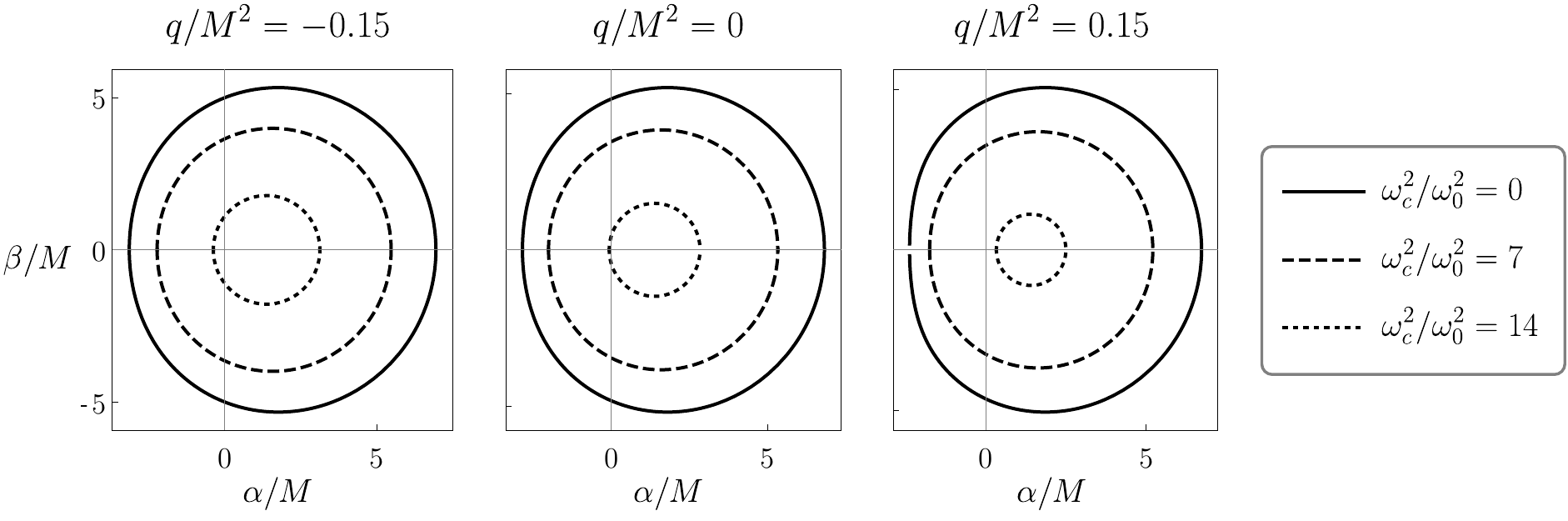}
	\caption{Shadow of a Kerr-Newman-like black hole with spin $a/M=0.9$ surrounded by a Shapiro-type plasma distribution with $f_r(r) = \omega_c^2 \sqrt{M^3 r}$, as viewed by an equatorial observer, for three values of the charge $q$ and three values of the photon frequency $\omega_0$.}
	\label{fig:sombras-KN}
\end{figure}

In our first example, we take the metric obtained by substituting into Eq. \eqref{eq:metrica} the function
\begin{equation}
	m(r) = M - \frac{q}{2r},
\end{equation}
with $M$ being a positive constant representing the mass of the black hole and $q$ an arbitrary real constant. If $q$ is positive and we set $q = Q^2$, the resulting metric corresponds to the Kerr-Newman spacetime with electric charge $Q$. However, we also allow negative values of $q$; such a metric arises in the absence of the electromagnetic field in alternative theories of gravity or within general relativity in the presence of certain matter fields. Horndeski gravity \cite{horndeski} is the most general scalar tensor theory with second-order derivative equations of motion, being the main theoretical framework for scalar-tensor models in which cosmological observations can be interpreted. Recently, the study of black holes has been of interest both in Horndeski and beyond Horndeski theories --with equations of motion of higher order in the derivatives, but with the property that the true propagating degrees of freedom obey well behaved second-order equations--. Special cases of Horndeski gravity admit a solution \cite{babichev17} having the same form as the Reissner-Nordstr\"om geometry in general relativity, but with the squared charge replaced by a constant that depends on parameters of the theory and can have any sign. Another interesting case arises within the Randall-Sundrum braneworld model \cite{rsbw}, in which the ordinary matter is in a three-dimensional space, the \textit{brane}, embedded in a larger space, the \textit{bulk}, where only gravity can propagate. The simplest (named ``second type'') of these theories consists of a positive tension brane in a bulk with only one extra dimension and a negative cosmological constant. The field equations, obtained with the help of the Gauss-Codazzi equations, admit a four-dimensional black hole solution on the brane \cite{aliev}, where $q$ is understood as a tidal charge generated by gravitational effects coming from a fifth dimension; for more details, see for example Ref. \cite{braneworld}, where the shadow without the presence of plasma is analyzed, and the references therein. For that reason and in analogy with the electric charge, in the following we will use the name ``charge'' for it. As in the Kerr-Newman case, the inequality $a^2 + q \leq M^2$ is a necessary and sufficient condition to avoid a naked singularity, and we assume that it is satisfied.

As described in Sec. \ref{shadow}, the black hole shadow can be obtained by plotting the parametric curve $(\alpha(r), \beta(r))$ that determines its contour, with $\alpha$ and $\beta$ given by Eqs. \eqref{eq:alfa} and \eqref{eq:beta}. Following Ref. \cite{perlick17}, it is straightforward to show that, for our chosen plasma density\footnote{The proof in fact works for any plasma density with $f_\theta \geq 0$ and $f_r = C r^k$, with $C \geq 0$ and $0 \leq k \leq 2$.}, the minus sign in Eqs. \eqref{eq:pphi-c} and \eqref{eq:q-c} is not physically realized, since the corresponding trajectories would have $\mathcal{K} < 0$, which can be shown to contradict the requirement that $\Theta \geq 0$. In addition, for the plasma model adopted here, a forbidden region appears if the frequency is low enough, where the condition of Eq. \eqref{eq:prohibida} is not satisfied. This axially symmetric region initially develops around the poles and expands towards the equator as the frequency decreases, eventually enveloping the black hole. 

\begin{figure}[t!]
	\centering
	\includegraphics[width=\textwidth]{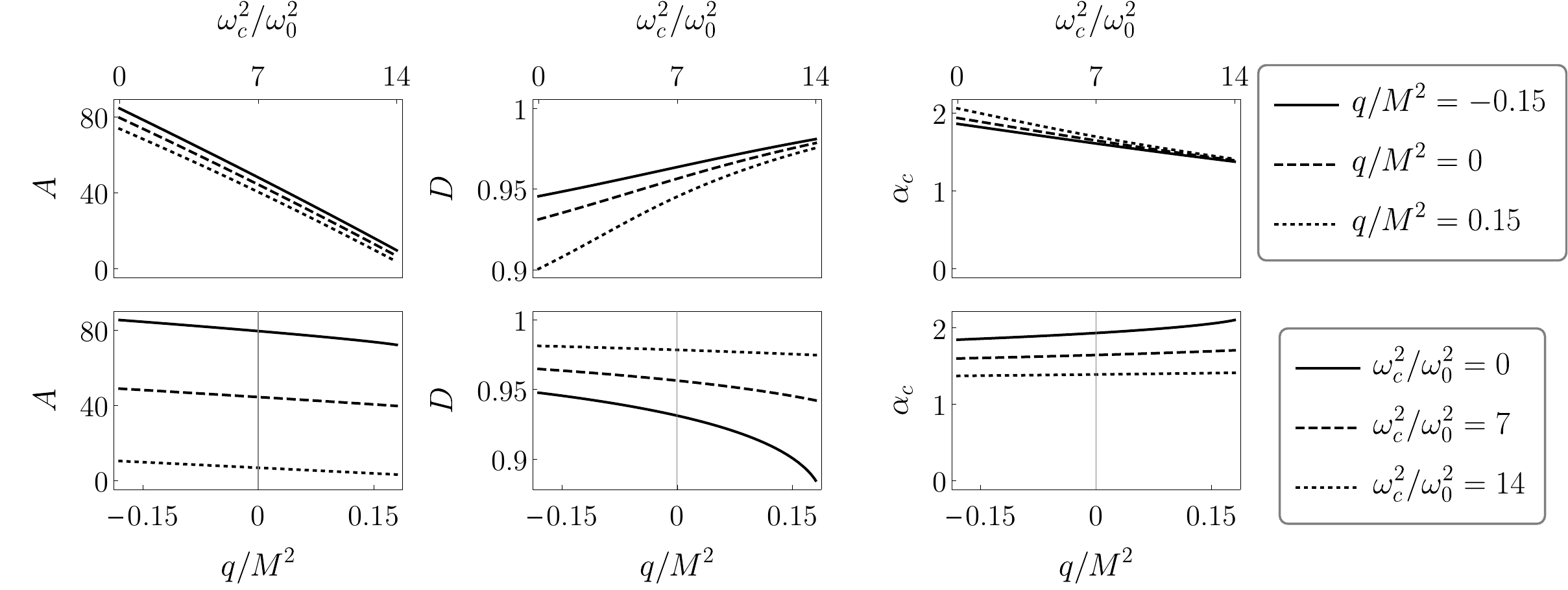}
	\caption{The area ($A$), the oblateness ($D$), and the centroid ($\alpha_c$) of the shadow of a Kerr-Newman-like black hole with spin $a/M = 0.9$ in a Shapiro-type plasma distribution with $f_r(r) = \omega_c^2 \sqrt{M^3 r}$, as viewed by an equatorial observer. \textit{Top}: the three observables as functions of the photon frequency $\omega_0$ for three values of the charge $q$. \textit{Bottom}: the three observables as functions of $q$ for three different values of $\omega_0$.}
	\label{fig:obs-KN}
\end{figure}

The contour of the shadow for a black hole with $a = 0.9M$ and some values for the photon frequency is displayed in Fig. \ref{fig:sombras-KN}, and the corresponding observables are shown in Fig. \ref{fig:obs-KN}. It is clear from the plots that the shadow becomes smaller and less deformed, and has the centroid closer to the origin as the photon frequency decreases; it disappears entirely below a certain frequency, due to the appearance of the forbidden region. We see that overall, the dependency of the observables on the frequency is stronger than on the charge $q$, except for the oblateness at small frequencies and near-extremal charge.

\subsection{Black holes in the scalar-tensor 4D Einstein-Gauss-Bonnet gravity}

\begin{figure}[t!]
	\centering
	\includegraphics[width=\textwidth]{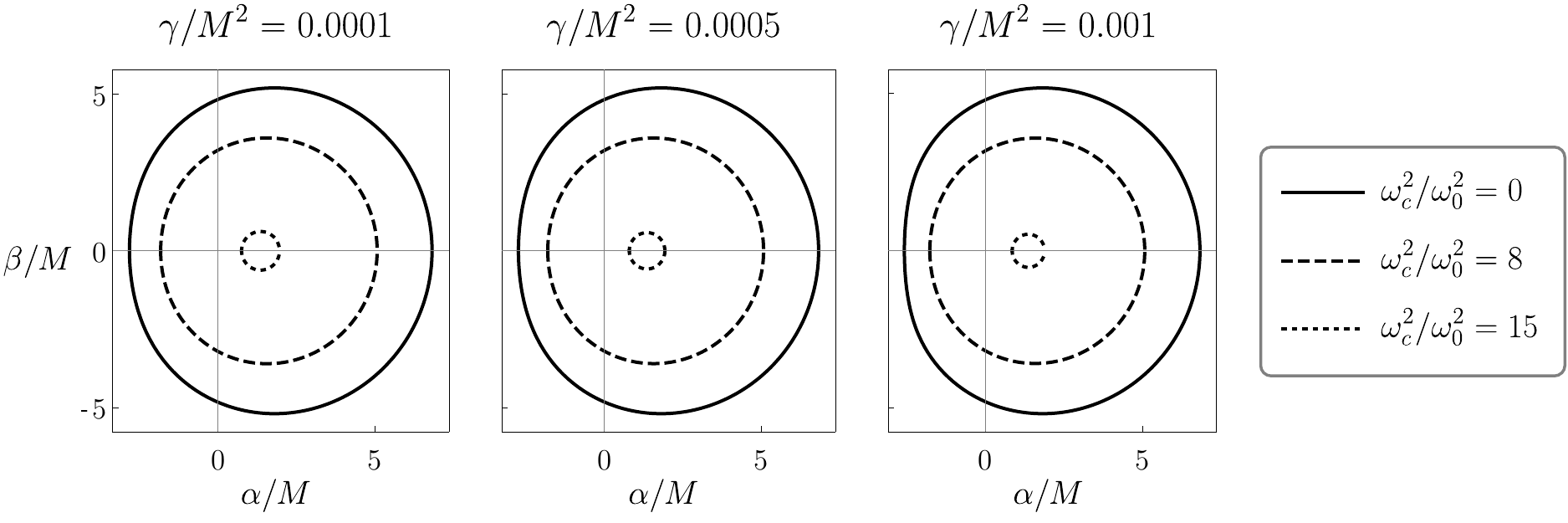}
	\caption{Shadow of a scalar-tensor 4D Einstein-Gauss-Bonnet black hole with spin $a/M=0.9$ surrounded by a Shapiro-type plasma distribution with $f_r(r) = \omega_c^2 \sqrt{M^3 r}$, as viewed by an equatorial observer, for three values of the parameter $\gamma$ and three values of the photon frequency $\omega_0$.}
	\label{fig:sombras-GB}
\end{figure}

Recently, a novel gravity model was proposed by rescaling the coupling constant and taking the limit $D \to 4$ in $D$-dimensional Einstein-Gauss-Bonnet theory \cite{glavan20}, with the purpose of bypassing the standard result that four-dimensional Einstein-Gauss-Bonnet theory is purely topological and thus equivalent to general relativity. However, this formulation is based on some particular solutions and it lacks of a complete set of well defined four-dimensional field equations \cite{gst}. Furthermore, it does not have an intrinsically four-dimensional description in terms of a covariantly conserved rank-2 tensor in four dimensions \cite{gst}. A proper theory should require additional scalar fields (of the Horndeski or Galileon type) to appear \cite{gst}. In order to solve these problems, a well-defined theory in four dimensions with a Gauss-Bonnet term was subsequently presented \cite{4degb}, which propagates a scalar field in addition to the metric tensor, and the full action belongs to the Horndeski class of scalar-tensor theories of gravity. The action is obtained by a regularization procedure in a way that is free from divergences and produces well behaved second-order field equations \cite{4degb}. The result is a new theory, dubbed scalar-tensor 4D Einstein-Gauss-Bonnet gravity, that includes a nonvanishing contribution coming from the Gauss-Bonnet term \cite{4degb}. Spherically symmetric black holes have been studied within this regularized theory \cite{4degb,4degb-bh}. A rotating solution was found \cite{kumar20} by applying a modified version of the Newman-Janis algorithm, resulting in a metric of the form shown in Eq.  \eqref{eq:metrica}, with the function $m(r)$ given by
\begin{equation}\label{eq:mEGB}
	m(r) = \frac{r^3}{64\pi \gamma} \left(\sqrt{1 + \frac{128\pi \gamma M}{r^3}} - 1\right),
\end{equation}
where $\gamma$ is a parameter of the theory, with units of mass squared; see Ref. \cite{clifton20} for a discussion of observational constraints on its value. The black hole mass is $M$, since we have $m(r) \to M$ as $r \to \infty$.  The limit $\gamma \to 0$ corresponds to  the Kerr geometry. The Newman-Janis algorithm does not guarantee \cite{hansen13} that the rotating metric will be a solution of the original field equations, so an appropriate set of field equations, possibly with an unknown matter component --transparent at the observed light frequency-- is assumed for this example. In general relativity, the energy-momentum tensor associated with the G\"urses-G\"ursey geometry has the form of an anisotropic fluid --isotropy is destroyed in the radial direction-- possibly describing a string fluid; for details, see Ref. \cite{gurses75} and references therein; in our case, it deserves a further study not necessary for our purposes. We will only consider positive values for $\gamma$, since the square root in Eq. \eqref{eq:mEGB} becomes imaginary for a finite value of $r$ if $\gamma$ is negative. Astrophysical constraints on compact objects within this theory have been recently presented \cite{charmousis21}.

\begin{figure}[t!]
	\centering
	\includegraphics[width=\textwidth]{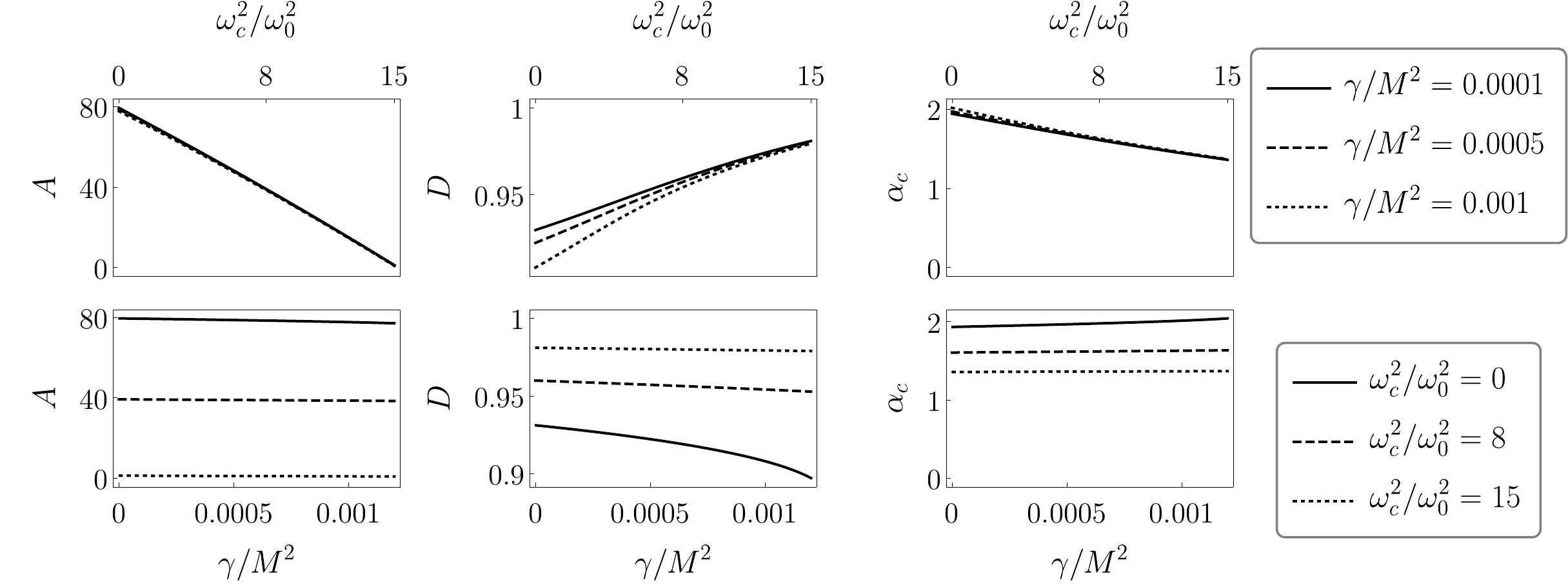}
	\caption{The area ($A$), the oblateness ($D$), and the centroid ($\alpha_c$) of the shadow of a scalar-tensor 4D Einstein-Gauss-Bonnet black hole with spin $a/M = 0.9$ in a Shapiro-type plasma distribution with $f_r(r) = \omega_c^2 \sqrt{M^3 r}$, as viewed by an equatorial observer. \textit{Top}: the three observables as functions of the photon frequency $\omega_0$ for three values of the parameter $\gamma$. \textit{Bottom}: the three observables as functions of $\gamma$ for three different values of $\omega_0$.}
	\label{fig:obs-GB}
\end{figure}

As in the previous subsection, we have plotted the black hole shadow for various values of the parameters, as shown in Figs. \ref{fig:sombras-GB} and \ref{fig:obs-GB}. We have again chosen a spin $a/M = 0.9$; by solving the equations $\Delta = \Delta' = 0$ numerically, it can be seen that the spacetime contains an event horizon if $\gamma/M^2 < 0.00129$, and there is a naked singularity for larger values of $\gamma$. Unlike in the previous example, it is not obvious in this case that the minus sign in Eqs. \eqref{eq:pphi-c} and \eqref{eq:q-c} is not relevant for the shadow; however, it can be verified numerically in this case that $\mathcal{K} < 0$ for the parameter values considered in this work. Similarly to the Kerr-Newman-like black hole, when including the Shapiro-type plasma distribution a forbidden region develops as the frequency of light decreases, resulting in a dramatic decrease of the shadow size, as seen in Fig. \ref{fig:sombras-GB}. This is also shown in Fig. \ref{fig:obs-GB}, which plots the three observables as functions of the parameters. The overall behavior of the observables is very similar to the one obtained for the Kerr-Newman-like black hole, with a smaller variation when changing the parameter, since the range of allowed values for $\gamma$ is much smaller than the one corresponding to the charge $q$.

\section{Conclusions}\label{conclu}

In this article, we have analyzed how the presence of plasma modifies the size and the shape of the shadow corresponding to a class of rotating black holes obtained by the Newman-Janis procedure. These spacetimes lead to a Hamilton-Jacobi equation for light rays that is always separable, as long as the plasma frequency $\omega_p$ satisfies the same condition previously adopted in the Kerr case \cite{perlick17}. The presence of plasma makes light follow timelike curves, leading to a modification of the photon regions and frequency-dependent forbidden regions, where light cannot travel. We have not considered the gravitational influence of the plasma itself, taking it to be negligible compared with that of the black hole, nor any processes of scattering, emission or absorption. These effects should be included if the goal is to produce a realistic image of the surroundings of a black hole and not only the shadow boundary.

Assuming that the plasma frequency obeys the separability condition introduced in Ref. \cite{perlick17}, we have obtained the expressions for the celestial coordinates of the shadow contour as viewed by a far away observer, which reduce to the already known ones \cite{tsukamoto18} when the plasma frequency is set to zero, or equivalently when the photon frequency tends to infinity. These expressions are the central result of this paper: for a metric obtained by the Newman-Janis algorithm and a plasma distribution satisfying the separability condition, one can plot the black hole shadow as seen by an observer at infinity, once given the values of the observer inclination and the photon frequency. One can also calculate various observables, such as the three we have defined: the area, oblateness and centroid of the shadow. These can be contrasted with observations of black hole shadows by the EHT or other future instruments, in order to place bounds on the parameters of alternatives to the Kerr metric. Among our observables, the displacement of the centroid is perhaps the most difficult one to determine observationally, since it requires an independent knowledge of the true position of the black hole in the sky. 

We have also applied our results to two example geometries, considering in both cases an equatorial observer for simplicity. For our plasma distribution, we have chosen a variation of the one proposed by Shapiro \cite{shapiro74}, which models dust surrounding a Kerr black hole and that is at rest at infinity. The original formulation is a function of $r$ only and thus does not obey the separability condition for the plasma density, so we have added a slight $\theta$ dependency to be able to use it within our formalism. Since we are not considering the Kerr metric but rather alternatives to it, especially at short distances, we also assume that the form of the corresponding plasma densities is not significantly altered from the original one, derived by assuming a Kerr black hole. As in the Kerr case, the presence of the plasma introduces a dependency on the photon frequency $\omega_0$ through the ratio $\omega_c/\omega_0$ (where $\omega_c$ is a constant) and thus this dependency is relative to the frequency scale set by $\omega_p$. In our first example, we have used a metric that we have named ``Kerr-Newman-like'', since it is identical in form to the Kerr-Newman metric but allows the parameter replacing the squared charge to take either sign. This metric arises in various scenarios involving matter fields, alternative theories such as Horndeski gravity, or possible effects of extra dimensions. The second example comes from a recently proposed theory, dubbed scalar-tensor four-dimensional Einstein-Gauss-Bonnet gravity, which has turned out to be of great interest. The shadows in both examples are qualitatively similar: for a given set of parameters of each model and fixed $\omega_c\neq 0$, the shadow becomes smaller and less deformed as $\omega_0$ decreases, with the clearest feature being the appearance of a forbidden region around the black hole. This forbidden region starts as two caps around the poles and grows towards the equatorial plane for decreasing $\omega_0$, which leads to a sharp reduction of the shadow size and its eventual disappearance. The presence of plasma always results in a smaller and less deformed shadow than in its absence ($\omega_c =0$). 

For the supermassive black holes at the centers of the Milky Way and the galaxy M87, which are the main focus of attention of current observational efforts by the EHT, it is expected that plasma effects start to become relevant at radio wavelengths of a few centimeters or more \cite{perlick15}. However, present and planned instruments focus on the submillimeter range, where scattering and self-absorption do not have a significant effect on the emitted radiation in the area surrounding the black hole, so that a realistic observation of the influence of a plasma on the shadow does not seem feasible at the moment.

\section*{Acknowledgments}

This work has been supported by CONICET and Universidad de Buenos Aires.

\end{document}